\documentclass
[floatfix,superscriptaddress,secnumarabic,amssymb,amsmath,nobibnotes,aps,prd,showkeys,showpacs,nofootinbib,onecolumn,notitlepage,12pt]{revtex4}%
\usepackage{setspace}
\usepackage{color}
\usepackage{amsmath}
\usepackage{amsfonts}
\usepackage{verbatim}
\usepackage{amssymb}
\usepackage{graphicx,bm}
\usepackage{graphicx}
\usepackage{amsmath}
\usepackage{amssymb}
\usepackage{amssymb}
\usepackage{graphicx,bm}
\usepackage{graphicx}
\usepackage[caption=false]{subfig}%
\providecommand{\U}[1]{\protect\rule{.1in}{.1in}}

\newcommand{\be}{\begin{equation}}
\newcommand{\ee}{\end{equation}}

\newcommand{\mincir}{\raise
-3.truept\hbox{\rlap{\hbox{$\sim$}}\raise4.truept\hbox{$<$}\ }}
\newcommand{\magcir}{\raise
-3.truept\hbox{\rlap{\hbox{$\sim$}}\raise4.truept\hbox{$>$}\ }}

\ifx\pdfoutput\relax\let\pdfoutput=\undefined\fi
\newcount\msipdfoutput
\ifx\pdfoutput\undefined\else
\ifcase\pdfoutput\else
\ifx\paperwidth\undefined\else
\ifdim\paperheight=0pt\relax\else\pdfpageheight\paperheight\fi
\ifdim\paperwidth=0pt\relax\else\pdfpagewidth\paperwidth\fi
\fi\fi\fi
\begin{document}
\title{The Equivalence Principle is NOT a Noether Symmetry}
\author{Andronikos Paliathanasis}
\email{anpaliat@phys.uoa.gr}
\affiliation{Institute of Systems Science, Durban University of Technology, Durban 4000,
South Africa}
\affiliation{Departamento de Matem\'{a}ticas, Universidad Cat\'{o}lica del Norte, Avda.
Angamos 0610, Casilla 1280 Antofagasta, Chile}

\begin{abstract}
The connection between the Equivalence Principle and Noether's theorem was
discussed in S. Capozziello and C. Ferrara, Int. J. Geom. Meth. Mod. Phys. 21,
2440014 (2024). However, it is known that the Noether symmetry condition is
independent of the equations of motions as follows from Hamilton's Principle.
In this paper, we critically examine the analysis presented in the
aforementioned work, highlighting its flaws, and provide a detailed
demonstration that there is no connection between Noether's theorem and the
Equivalence Principle. Furthermore, we offer various insights on symmetry
analysis to prevent the perpetuation of inaccuracies regarding Noether's work.

\end{abstract}
\maketitle

\section{Introduction}

\label{sec1}

Symmetry analysis plays a critical role in the analytical treatment of
nonlinear differential equations. Symmetries are essential in many aspects of
physical science, particularly in gravitational physics, and are thoroughly
explored in the literature. However, despite the extensive number of studies
in mathematical science and physics on the treatment of symmetries, many
inaccuracies persist in the literature, leading to incorrect conclusions and results.

In a recent study \cite{cap1}, they investigated the relation of Noether's
theorem with that of the Einstein's Equivalence Principle. The authors
concluded that Einstein's Equivalence Principle is a Noether symmetry for
General Relativity and for metric-affine theories. However, these conclusions
were reached due to a flawed application of symmetry analysis and a
misinterpretation of Noether's theorems. In this study, we critically examine
the methodology used in \cite{cap1} and demonstrate that their results are
incorrect. We firmly state that Einstein's Equivalence Principle is not a
Noether symmetry. The structure of the paper is organized as follows.

In Section \ref{sec3} we discuss the general form of Noether's theorems. In
Section \ref{sec4} we present the application of Noether's symmetry conditions
for the geodesic Lagrangian, where we point the flawed results of \cite{cap1}.
Finally, in Section \ref{sec5} we draw our conclusions.

\section{Noether's theorems}

\label{sec3}

Symmetry plays a fundamental role in numerous aspects of the physical world.
The systematic treatment of symmetry analysis was introduced in the late 19th
century by Sophus Lie, whose work revolutionized the approach to differential
equations. Lie's contributions introduced the concept of considering the
infinitesimal representations of finite transformations of continuous groups.
This methodology facilitated the transition from group theory to a local
algebraic representation, enabling a deeper exploration of invariance
properties under these transformations.

The primary goal of determining the invariant transformations that leave a
given differential equation unchanged is to simplify the solution of the
equation under study. The presence of symmetries enables one to solve
differential equations through repeated reduction of order, often using a
reverse series of quadratures or by determining a sufficient number of first integrals.

A few decades later based on the spirit of Lie's approach, Emmy Noether
presented her pioneer work which include two main theorems \cite{noether18}.
Noether's first theorem treats the invariance of the functional of the
Calculus of Variations -- the Action Integral in Mechanics -- under an
infinitesimal transformation. On the other hand, Noether's second theorem
relates variational symmetries to conserved quantities. Notably, Noether's
work allows the coefficient functions of the infinitesimal transformations to
depend on the derivatives of the dependent variables, extending the
applicability of symmetry analysis.

Moreover, the boundary function on the Action Integral can include
higher-order derivatives of the dependent variables. Thus, a series of studies
which discuss generalizations of Noether's work are all included on Emmy
Noether's original work, for more details we refer the reader to the recent
discussion in \cite{noeleach}. Some prior studies of Noether's work which
investigate the symmetries are the independent studies of Hamel
\cite{Hamel59,Hamel50}, of Herglotz \cite{Herglotz36}, of Knesner
\cite{Knesner} and of Klein \cite{Klein}. An English translation of Noether's
work and more historical details on the foundations of the two theorems are
presented in \cite{bknoe}.

In the following lines we present the basic elements of Noether's theorems.

Let us introduce the infinitesimal transformation%
\begin{equation}
\bar{t}=t+\varepsilon\xi,\qquad\bar{q}=q+\varepsilon\eta, \label{it.01}%
\end{equation}
where the variable dependences of the functions $\xi$ and $\eta$ is arbitrary
and $\varepsilon$ is an infinitesimal parameter, that is, $\varepsilon
^{2}\rightarrow0$.

We consider the Action Integral%
\begin{equation}
S=\int_{t_{0}}^{t_{1}}L\left(  t,q,\dot{q}\right)  dt. \label{it.02}%
\end{equation}
where $L\left(  t,q,\dot{q}\right)  $ is the Lagrange function. Under the
Action of the infinitesimal transformation (\ref{it.01}) reads%
\begin{equation}
\bar{S}=\int_{\bar{t}_{0}}^{\bar{t}_{1}}L\left(  \bar{t},\bar{q},\dot{\bar{q}%
}\right)  d\bar{t}. \label{it.03}%
\end{equation}
Thus, expanding in first order of the parameter $\varepsilon$ we find%
\begin{align}
\bar{A}  &  =\int_{t_{0}}^{t_{1}}\left[  L+\varepsilon\left(  \xi
{\frac{\partial L}{\partial t}}+\eta{\frac{\partial L}{\partial q}}%
+\zeta{\frac{\partial L}{\partial\dot{q}}}+\dot{\xi}L\right)  \right]
dt\nonumber\\
&  +\varepsilon\left[  \xi{t_{1}}L(t_{1},q_{1},\dot{q}_{1})-\xi{t_{0}}%
L(t_{0},q_{0},\dot{q}_{0})\right]  , \label{it.04}%
\end{align}
or equivalently%
\begin{equation}
\bar{A}=A+\varepsilon\int_{t_{0}}^{t_{1}}\left(  \xi{\frac{\partial
L}{\partial t}}+\eta{\frac{\partial L}{\partial q}}+\zeta{\frac{\partial
L}{\partial\dot{q}}}+\dot{\xi}L\right)  dt+\varepsilon F, \label{it.05}%
\end{equation}
where
\begin{equation}
F=\xi{t_{1}}L(t_{1},q_{1},\dot{q}_{1})-\xi{t_{0}}L(t_{0},q_{0},\dot{q}_{0}),
\end{equation}
$L(t_{0},q_{0},\dot{q}_{0})$ and $L(t_{1},q_{1},\dot{q}_{1})$ are the values
of $L$ at the endpoints $t_{0}$ and $t_{1}$ respectively, and
\begin{equation}
\zeta=\dot{\eta}-\dot{q}\dot{\tau}.
\end{equation}

Since function $F$ depends only upon the endpoints it can be written as
$F=-\int_{t_{0}}^{t_{1}}\dot{f}dt.$ Therefore, the variation of the Action
Integral (\ref{it.02}) remain invariant under the application of the
infinitesimal transformation (\ref{it.01}) if and only if%
\begin{equation}
\xi{\frac{\partial L}{\partial t}}+\eta{\frac{\partial L}{\partial q}}%
+\zeta{\frac{\partial L}{\partial\dot{q}}}+\dot{\xi}L-\dot{f}=0
\end{equation}
that is%
\begin{equation}
X^{\left[  1\right]  }L+\dot{\xi}L-\dot{f}=0, \label{it.06}%
\end{equation}
in which $X^{\left[  1\right]  }=X+\zeta\partial_{\dot{q}}$ is the first
extension of the vector field $X$ in the jet space, and
\begin{equation}
X=\xi\partial_{t}+\eta\partial_{q}.
\end{equation}
is the generator of the infinitesimal transformation (\ref{it.01}). When
condition (\ref{it.06}) the vector field is called Noether symmetry. Without
loss of generality the coefficients of the vector field $X$, i.e. $\xi$ and
$\eta$ is arbitrary which means that they can depend on the derivatives of the
dependent variables or on more general forms.

Therefore, the classification of Noether's symmetries as\ presented in
\cite{cap1} and \cite{cap2} into Generalized Noether symmetries, when
$\dot{\xi}\neq0$, Noether symmetries for canonical Lagrangians $\dot{\xi
}=0,~\dot{f}=const$, and internal Noether symmetries $\dot{\xi}=0,~\dot{f}=0$,
it is at least unnecessary. It can lead to non mathematically acceptable
conclusions and makes the issue of symmetries problematic. It shows lack on
the mathematical literature of the subject of symmetry.

If we impose Hamilton's principle for the Action Integral (\ref{it.02}) we end
up with the Euler-Lagrange equations of motion%
\begin{equation}
\frac{d}{dt}\left(  \frac{\partial L}{\partial\dot{q}}\right)  -\frac{\partial
L}{\partial q}=0. \label{it.07}%
\end{equation}

Hence, with the use of the equations of motion (\ref{it.07}) the Noether
symmetry condition (\ref{it.06}) can be written in the equivalent form%
\begin{equation}
\frac{d}{dt}\left(  f+\xi\mathcal{H}-\eta p\right)  =0, \label{it.08}%
\end{equation}
in which $\mathcal{H}=p\dot{q}-L$ is the Hamiltonian function and
$p=\frac{\partial L}{\partial\dot{q}}$ is the momentum. The latter expression
states that the function%
\begin{equation}
\Phi\left(  t,q,p\right)  =\xi\mathcal{H}-\eta p+f,
\end{equation}
is a conserved quantity for the equations of motion (\ref{it.07}). This is
known as Noether's second theorem.

We emphasize that symmetry is the generator of an infinitesimal transformation
that leaves the Action Integral invariant, and the existence of such symmetry
is independent of the Euler-Lagrange Equation from the Calculus of Variations.
The Euler-Lagrange equation arises from the application of Hamilton's
Principle, where the variable $q$ is subjected to a zero-endpoint variation.
However, no such restriction applies to the infinitesimal transformations
introduced by Noether. Indeed the equation of motion can exist either if the
Action Integral does not possess any symmetries.

This observation clearly states that there can be no connection between
the\ existence of variational symmetries and the existence for the equations
of motion, or the equivalence principle.\ Nevertheless, in the following lines
we will show in a pedagogical way how the authors of \cite{cap1} were driven
to a faulty conclusion.\hfill

\section{Noether symmetries for Geodesic Lagrangian}

\label{sec4}

Consider a second-rank tensor $g_{\mu\nu}\left(  q^{k}\right)  $ which is a
metric tensor, and the Lagrangian function%
\begin{equation}
L\left(  t,q,\dot{q}\right)  =\frac{1}{2}g_{\mu\nu}\dot{q}^{\mu}\dot{q}^{\nu}.
\label{it.09}%
\end{equation}
We follow \cite{cap1} and we study the Noether symmetry condition
(\ref{it.06}) for the Lagrangian function (\ref{it.09}). As in \cite{cap1} we
focus in the case where the symmetry vector is $X=\eta\partial_{q}$ and we
investigate for transformations where the boundary term is zero, i.e. $\dot
{f}=0$.

Indeed, for this consideration the symmetry condition (\ref{it.06}) reads%
\begin{equation}
X^{\left[  1\right]  }L=0. \label{it.10}%
\end{equation}
where now $X^{\left[  1\right]  }=\eta\partial_{q}+\dot{\eta}\partial_{\dot
{q}}$. \ 

In \cite{cap1} it has been considered that the symmetry condition is $X\nabla
L=0$, where$~\nabla$ describes the covariant derivative. However, the correct
expansion of condition (\ref{it.10}) which can be found in any standard
textbook of symmetry analysis
\begin{equation}
X^{\left[  1\right]  }L\equiv\mathcal{L}_{X^{\left[  1\right]  }}L=L_{,q}%
\eta+L_{,\dot{q}}\dot{\eta}.
\end{equation}
This indeed is a (guided) derivative in the jet space $\left\{  q,\dot
{q}\right\}  $, but it should not be confused with the covariant derivative
$\nabla~$of the background geometry. This observation is sufficient to
conclude that the main results of the analysis in \cite{cap1} are at least
inaccurate, and with the discussion we made before we can conclude that the
equivalent principle is NOT a Noether symmetry.

Another point which deserve discussion is that Noether symmetries for a given
Lagrangian leads to conservation laws for the Euler-Lagrange equations of the
studied Lagrangian. Specifically, the application of Hamilton's principle for
the Lagrangian (\ref{it.09}) leads to the equations of motion%
\begin{equation}
\ddot{q}^{\mu}+\mathring{\Gamma}_{\kappa\nu}^{\mu}\dot{q}^{\kappa}\dot{q}%
^{\nu}=0,
\end{equation}
where $\mathring{\Gamma}_{\kappa\nu}^{\mu}$ is the Levi-Civita connection
which defines the covariant derivative $\mathring{\nabla}$ with the property
$\mathring{\nabla}_{\kappa}g_{\mu\nu}=0$. However, in \cite{cap1} the authors
considered the autoparallels which depend on a more general connection, and it
is not clear how the geodesic\ Lagrangian (\ref{it.09}) is related to the
autoparallels for an arbitrary connection.

Furthermore, in \cite{cap1} the authors explore the relation between the
Noether symmetries for the geodesic Lagrangian with the symmetries of the
metric tensor $g_{\mu\nu}$. Before concluding this work, for the sake of
clarity, we note that in \cite{nsg} the geometric nature of the Noether
symmetries was investigated when the coefficients $\xi$ and $\eta$ are
functions of $t$ and $q$. Specifically, the Noether symmetries for the
geodesic Lagrangian (\ref{it.09}) are generated by the elements of the
Homothetic group of the metric tensor $g_{\mu\nu}$ and form a subgroup which
belongs to the special projective group for the $n+1$ decomposable spacetime
as discussed in details in \cite{ns2}. On the hand, when $\xi$ and $\eta$ are
linear on the derivative $\dot{q}$, the symmetry conditions indicate that the
generator of the symmetry vectors are the Killing tensors of the metric tensor
$g_{\mu\nu}~$\cite{sarlet}. See also the discussion in \cite{ts} and
references therein.

\section{Conclusions}

\label{sec5}

In this study, we provide a detailed discussion of Noether's theorem and its
relationship with the equations of motion. Specifically, with regard to the
geodesic Lagrangian, we demonstrate that there is no connection between the
Equivalence Principle and Noether's symmetries.

Finally, we made various clarifications on what is called as a Noether
symmetry, and we mentioned the complete connection of the Noether symmetries
with the collineations of the metric tensor $g_{\mu\nu}$. \ Hopefully this
work it helps to avoid perpetuating various inaccuracies on Noether's work.

\textbf{Data Availability Statements:} Data sharing is not applicable to this
article as no datasets were generated or analyzed during the current study.

\textbf{Code Availability Statements:} Code sharing is not applicable to this
article as no code was used in this study.

\begin{acknowledgments}
AP thanks the support of VRIDT through Resoluci\'{o}n VRIDT No. 096/2022 and
Resoluci\'{o}n VRIDT No. 098/2022. Part of this study was supported by
FONDECYT 1240514.
\end{acknowledgments}


\begin{thebibliography}{99}                                                                                               %


\bibitem {cap1}S. Capozziello and C. Ferrara, Int. J. Geom. Meth. Mod. Phys.
21, 2440014 (2024)

\bibitem {cap2}F. Bajardia and S.\ Capozziello, Noether Symmetries in Theories
of Gravity, Cambridge University Press,\ Cambridge (2022)

\bibitem {noether18}E. Noether, Invariante Variationsprobleme, K\"{o}niglich
Gesellschaft der Wissenschaften G\"{o}ttingen Nachrichten Mathematik-physik
Klasse 2, 235 (1918)

\bibitem {noeleach}A. Halder, A. Paliathanasis and P.G.L. Leach, Symmetry 10,
744 (2018)

\bibitem {Hamel59}G. Hamel, Ueber die Grundlagen der Mechanik, Mathematische
Annalen 66, 350 (1908)

\bibitem {Hamel50}G. Hamel, Ueber ein Prinzig der Befreiung bei Lagrange,
Jahresbericht der Deutschen Mathematiker-Vereinigung 25, 60 (1917)

\bibitem {Herglotz36}G. Herglotz, \"{U}ber den vom Standpunkt des
Relativit\"{a}tsprinzips aus als starr\textquotedblright\ zu bezeichnenden
K\"{o}rper, Annalen der Physik 336, 393 (1910)

\bibitem {Knesner}A. Kneser, Kleinste Wirkung und Galileische Relativit\"{a}t.
Mathematische Zeitschrift 2 326 (1918)

\bibitem {Klein}F. Klein, K\"{o}niglich Gesellschaft der Wissenschaften,
G\"{o}ttingen Nachrichten Mathematik-physik Klasse 2, (1918)

\bibitem {bknoe}Y. Kosmann-Schwarzbach and B.E. Schwarzbach, The Noether
Theorems: Invariance and Conservation Laws in the Twentieth Century, Springer
New York, New York (2011)

\bibitem {nsg}M. Tsamparlis and A. Paliathanasis, Gen. Relat. Gravit. 43, 1861 (2011)

\bibitem {ns2}A. Paliathanasis, Symmetry 13, 1068 (2021)

\bibitem {sarlet}W. Sarlet and F. Cantrijin, J. Phys.\ A: Math. Gen. 16, 1383 (1983)

\bibitem {ts}A. Mitsopoulos and M.\ Tsamparlis, J. Math. Phys. 64, 012701 (2023)
\end{thebibliography}
\end{document}